\documentclass[fleqn]{vch-book}
\usepackage[dvips]{color}
\usepackage{graphicx}
\usepackage{amsmath}
\usepackage{amssymb}

\newcommand{\mbb}[1]{\mbox{\boldmath $#1$}}
\newcommand{\mg}{\mbb{\Gamma}}
\newcommand{\openone}{\textbf{1}}

\begin{document}


\chapter{Non-classical Gaussian states in noisy environments}

\tocauthor{Stefan Scheel and Dirk--Gunnar Welsch}
\authorafterheading{Stefan Scheel\,%
\footnote{Quantum Optics and Laser Science, Blackett Laboratory,
Imperial College, Prince Consort Road, London SW7 2BW, UK, email:
\tt{s.scheel@ic.ac.uk}}
and Dirk--Gunnar Welsch\,%
\footnote{Theoretisch-Physikalisches Institut,
Friedrich-Schiller-Universit\"at Jena, Max-Wien-Platz 1, D-07743 Jena,
Germany, email: \tt{welsch@tpi.uni-jena.de}}
}

\section{Introduction}
\label{sec:intro}

Research in the field of quantum information theory nowadays focusses
more and more on the possibilities of practical implementation of
quantum information processes. One of the most promising attempts
has seemed to be the usage of entangled Gaussian states since they are
rather easy to produce with optical means. In this article we will
review properties of Gaussian states and describe operations on
them. The interaction of the electromagnetic field with an absorbing
dielectric as a special type of environmental interaction will serve
as the basis for the understanding of decoherence and entanglement
degradation of Gaussian states of light propagating through fibers.

In Section \ref{sec:gauss} we shortly review the definition of Gaussian
states and operations that can be performed on them. These includes
symplectic (unitary) transformations, trace-preserving maps, and projective
measurements. The important definitions of classicality, separability, and
entanglement are defined and appropriate measures are given for them.
In Section \ref{sec:degradation} we review some results on entanglement
degradation in dielectric environments. We discuss the decoherence of
Gaussian quantum states of light transmitted through absorbing
dielectric objects such as fibers.
Section \ref{sec:teleport} is devoted to the study of quantum
teleportation in noisy environments. Special emphasis
is put onto the question of choosing the correct displacement on the
receiver's side.


\section{Gaussian states and Gaussian operations}
\label{sec:gauss}

To begin with, we define Gaussian states and Gaussian operations
and discuss some important properties such as non-classicality, separability,
and entanglement.
A quantum state is called Gaussian if its
characteristic function, and hence an appropriately chosen
phase-space function, is an
exponential form that is at most quadratic in its canonical variables.
Most importantly, a Gaussian state is fully characterized
by the first and second moments of the $2N$ canonical variables
$\,\hat{\!\mbb{\xi}}{^{\rm T}}$ $\!=$
$\!(\hat{x}_1,\hat{p}_1,\ldots,\hat{x}_N,\hat{p}_N)$
obeying the canonical commutation relations ($\hbar$ $\!=$ $\!1$)
\begin{equation}
\label{eq:2.2}
\left[ \hat{\xi}_i, \hat{\xi}_j \right] = i\Sigma_{ij}, 
\end{equation}
where $\mbb{\Sigma}$ is the $2N$-dimensional symplectic matrix
\begin{equation}
\mbb{\Sigma}
=\left(
\begin{array}{cc}0&1\\-1&0\end{array}
\right)^{\oplus N}.
\end{equation}
This definition seems to suggest that higher moments do not play any
r\^{o}le. However, this is not correct, since all higher moments can
be derived from first and second moments. The first moments describe
the position $\mbb{\kappa}$ of the `centre-of-mass' of the Gaussian
function in phase-space, whereas the second moments describe the
fluctuations of the canonical variables or equivalently, the width of
the Gaussian in phase-space. They can be collected in the positive
symmetric $2N\times 2N$-dimensional covariance matrix $\mg$, viz.
\begin{equation}
\Gamma_{ij}=2\mbox{Tr}\!\left[  
\bigl(\hat{\xi}_i-\langle \hat{\xi}_i\rangle\bigr)
\bigl(\hat{\xi}_j-\langle \hat{\xi}_j\rangle\bigr)
\hat{\varrho} \right]-i\Sigma_{ij} .
\end{equation}
The associated charateristic function is the Fourier transform of the
Wigner function. Note that without the term $-i\Sigma_{ij}$ one would
obtain the covariance matrix corresponding to the density operator in
normal order.
The fluctuations, however, are not independent
from each other. This means that not every positive $2N\times 2N$-dimensional
matrix is an admissible covariance matrix, i.e. describes a physical state.
The reason for this behaviour is the simple fact that the fluctuations have
to obey the Heisenberg uncertainty relations which in matrix form read as
\begin{equation}
\label{eq:2.4}
\mg+i\mbb{\Sigma} \ge 0 .
\end{equation}
With the above definitions we can write the characteristic function of
a general Gaussian state as
\begin{equation}
\label{eq:2.5}
\chi(\mbb{\lambda}) =
\exp\left[-{\textstyle\frac{1}{4}}\mbb{\lambda}^{\rm T}\mg\mbb{\lambda}
+i\mbb{\lambda}^{\rm T}\mbb{\Sigma}\mbb{\kappa} \right] .
\end{equation}

For the following discussion we will focus on the second moments only.
We can transform Gaussian states into each other by transforming the
covariance matrix as
\mbox{$\mg$ $\!\to$ $\!\mg'$ $\!=$ $\!\textbf{S}\mg\textbf{S}^{\rm T}$},
where the matrix $\textbf{S}$ has to be chosen such that the
canonical commutation relations (\ref{eq:2.2})
are fulfilled. This requirement
leads to the constraint \mbox{$\textbf{S}\mbb{\Sigma}\textbf{S}^{\rm T}$
$\!=$ $\!\mbb{\Sigma}$}.
On the level of states, these matrices generate unitary transformations
$\hat{\varrho}$ $\!\to$ $\!\hat{\varrho}'$
$\!=$ $\!\hat{U}(\textbf{S})\hat{\varrho}\hat{U}^\dagger(\textbf{S})$.
In turn, any unitary transformation which is bilinear in the canonical
variables can be described by a symplectic matrix.
All symplectic matrices $\textbf{S}$ form the 
(non-compact) symplectic group Sp($2N$,$\mathbb{R}$). As for every group,
it has certain normal forms associated with it. One of the most important
ones is the generalized Euler decomposition
\cite{Arvind95}:
Every symplectic matrix $\textbf{S}$ can be decomposed into three factors,
$\textbf{S}=\textbf{O}_1\textbf{D} \textbf{O}_2$,
where the $\textbf{O}_i$ belong to the $N$-dimensional (compact) orthogonal
group and $\textbf{D}$ is a diagonal matrix with entries 
$(k_1,1/k_1,\ldots,k_N,1/k_N)$. This decomposition has an immediate physical
interpretation: The $\textbf{O}_i$ generate $N$-mode rotations (generated by
beam splitter networks and phase shifters) and the matrix $\textbf{D}$ 
describes single-mode squeezings.


\subsection{Classicality}

A very special feature of a quantum state is its possible non-classical
character, which means that its inherent statistics cannot be modelled by a
classical probability distribution. A widely used definition has been
given by Titulaer and Glauber \cite{Titulaer65}
saying that a state is classical (or `predetermined' in their words)
if and only if its $P$ function is positive semi-definite and
sufficiently smooth, hence $P\in L_2(\mathbb{R}^+)$. This definition
has recently been translated into a measurable criterion which is
based on quadrature distributions \cite{Vogel00}.

Let us discuss the implications for a general Gaussian $N$-mode
state. The main result is that we regard a Gaussian state as
non-classical whenever one or more eigenvalues of its covariance matrix
drop below $1$. The reasoning behind the argument is that in such a
case the characteristic function does not possess a Fourier transform
since it increases exponentially at infinity.
To answer the question of whether a Gaussian state is classical or
not, we recall that any
Gaussian state (with zero mean) can be generated by acting on
a thermal state (being classical) with a symplectic matrix
$\textbf{S}$. That is, we can
start with a covariance matrix of the form
\mbox{$\mg_{\rm th}$ $\!=$ $\!2 \mbox{diag}(n_1,n_1,\ldots,n_N,n_N)$
$\!+$ $\!\openone$},
with $n_i$ being the mean thermal excitation number in the $i$-th mode.
Next we act on $\mg_{\rm th}$ with a symplectic matrix $\textbf{S}$
and compute the eigenvalues of the transformed covariance matrix
$\textbf{S}\mg_{\rm th}\textbf{S}^{\rm T}$ with \mbox{$\textbf{S}$ $\!=$
$\!\textbf{O}_1\textbf{D} \textbf{O}_2$}.

To give a simple example, let us consider a single-mode Gaussian state.
The sought eigenvalues are simply the eigenvalues of the covariance matrix
($k$ $\!=$ $\!e^\zeta$)
\begin{equation}
\textbf{D}\mg_{\rm th}\textbf{D}
=\textrm{diag}\left[e^{2\zeta}(2n+1),e^{-2\zeta}(2n+1)\right]
\end{equation}
(the orthogonal matrices $\textbf{O}_i$ do not play a r\^{o}le here),
from which we immediately find the well-known relation 
$|\zeta_{\max}|$ $\!=$ $\!\frac{1}{2} \ln (2n+1)$
for the maximal squeezing value such that the state stays classically
\cite{Marian93}.
We see that a state can be classical, though it is squeezed. 
Needless to say that multimode Gaussian states can be treated
in the same way. Clearly, the number of parameters one can choose
increases rapidly.

Furthermore, a non-classicality
measure $C(\hat{\varrho})$ based
on the relative entropy can be defined as
\begin{equation}
C(\hat{\varrho}) = \min_{\hat{\sigma}_{\rm class.}}
\mbox{Tr} \left[ \hat{\varrho} \left( \ln \hat{\varrho} -
\ln \hat{\sigma} \right) \right].
\end{equation}
The advantage of using this entropic measure for Gaussian states is
that it can be easily calculated \cite{Scheel01}. Note that this 
measure can be equally well used as the measure based on
Bures' metric defined in \cite{Marian02}.


\subsection{CP maps and partial measurements}
\label{sec:cpmap}

Until now we have described unitary operations on the level of states.
Decoherence processes, however, correspond to operations in a larger
class, the completely positive (CP) maps. Physically, they are unitary
operations in a larger Hilbert space (Naimark theorem \cite{Naimark}).
A typical example is the interaction with a dissipative
environment which is traced out afterwards. On
the level of covariance matrices, CP maps act as
\begin{equation}
\label{eq:2.7}
\mg \to
\mg' =
\mbb{A}\mg\mbb{A}^{\rm T} +\mbb{G},
\end{equation}
where $\mbb{G}$ is a positive symmetric matrix and $\mbb{A}$
an arbitrary matrix, with the only constraint being
that the resulting matrix $\mg'$ should be a valid
covariance matrix, i.e. fulfils Eq.~(\ref{eq:2.4}). The
non-orthogonality of the matrix $\mbb{A}$
is a direct consequence of dissipation, and $\mbb{G}$ is the
additional noise as required by the dissipation-fluctuation theorem.
As a simple example, $\mbb{G}$ can represent pure thermal noise. In
this case, one can produce a thermal state with covariance matrix  
$\mg_{\rm th}$ $\!=$ $\!(2n$ $\!+$ $\!1)\openone$ from the vacuum with
$\mg_{\rm vac}$ $\!=$ $\!\openone$ with the help of
the matrices $\mbb{A}$ $\!=$ $\!\openone$ and $\mbb{G}$ $\!=$
$\!2n\openone$. 

Another important class of operations is generated by
(homodyne) measurements on subsystems.
For that, let us write the covariance matrix
of a bipartite system in block form
$\mg$ $\!=$
$\!\left(\begin{array}{cc}\textbf{C}_1&\textbf{C}_3\\
\textbf{C}_3^{\rm T}&\textbf{C}_2\end{array}\right)$
where $\textbf{C}_1$ and $\textbf{C}_2$ are the principal submatrices with
respect to the subsystems $1$ and $2$, respectively, and
$\textbf{C}_3$
is the correlation matrix between the subsystems. In general,
the dimension of all
these matrices can be different, e.g., $\textbf{C}_1$ is $2N\times 2N$, 
$\textbf{C}_2$ is $2M\times 2M$, and $\textbf{C}_3$ is $2N \times 2M$.
Projection  measurements on subsystem $2$ onto a Gaussian state with
covariance matrix $\textbf{D}$ leads to the Schur complement \cite{Eisert02} 
with respect to the leading principal submatrix $\textbf{C}_1$,
\begin{equation}
\label{eq:2.8}
\mg \to
\mg' =
\textbf{C}_1-\textbf{C}_3 \left( \textbf{C}_2+\textbf{D}^2
\right)^{-1} \textbf{C}_3^{\rm T}.
\end{equation}
The probability of measuring a certain outcome is
\mbox{$\bigl[\det\,\big(\textbf{C}_2$ $\!+$
$\!\textbf{D}^2\big)\bigr]^{-1/2}$}.
Note that the dimension of the matrix $\mg'$ is now reduced.
When $\textbf{D}$ is the identity matrix, then the
transformation (\ref{eq:2.8}) describes
vacuum projection. For 
$\textbf{D}$ $\!=$ $\!\lim_{d\to 0}(1/d,d,\ldots,1/d,d)$ it describes
homodyne detection, in which case the inverse has to be thought of as the
Moore--Penrose (MP) inverse \cite{Horn}. An analogous treatment has to
be made for the corresponding determinant.
Combining the transformations (\ref{eq:2.7}) and (\ref{eq:2.8})
gives the most general
(not necessarily trace-preserving) Gaussian operation where the name means
that they leave the Gaussian character of a state invariant.


\subsection{Separability and entanglement}
\label{sec:sepent}

So far, we have discussed general features of Gaussian states and Gaussian
operations. In what follows, we will focus on bipartite Gaussian states
and their entanglement properties. To be more precise, we restrict ourselves
to the class of ($1\times 1$)-mode states, in which one mode of the
electromagnetic field is entangled with another one.
The covariance matrix $\mg$ is thus a ($4\times 4$)-matrix.
In order to check whether a given bipartite Gaussian state is entangled or
not, a necessary and sufficient separability criterion has been developed 
\cite{Duan00,Simon00}. It is equivalent to the Peres--Horodecki criterion
for entangled qubits \cite{PeresHorodecki}. The test consists of
checking whether the \textit{partial} transpose of the density operator
is again a valid density operator. In this case, the state is separable
and therefore not entangled. On the level of the covariance matrix, the
partial transpose consists of pre- and post-multiplying $\mg$ by
the diagonal matrix \mbox{$\textbf{P}$ $\!=$ $\!\mbox{diag}(1,1,1,-1)$},
viz.
$\mg^{\rm P.T.}$ $\!=$ $\!\textbf{P}\mg\textbf{P}^{\rm T}$.
Hence, if $\mg^{\rm P.T.}$ $\!+$ $\!i\mbb{\Sigma}$ $\!\ge$ $\!0$
the state is separable. In the block notation used in Section
\ref{sec:cpmap}, this criterion translates as \cite{Simon00}
\begin{equation}
\label{eq:sep}
\hspace*{-6ex}
\det\textbf{C}_1 \textbf{C}_2 
+\big( 1 -|\det\textbf{C}_3| \big)^2
-\mbox{Tr}\left[ \textbf{C}_1 \mbb{\Sigma} \textbf{C}_3 \mbb{\Sigma}
\textbf{C}_2 \mbb{\Sigma} \textbf{C}_3^{\rm T} \mbb{\Sigma} \right] 
\ge \det\textbf{C}_1 + \det\textbf{C}_2 \,.
\end{equation}

Once one has checked for inseparability, one may ask
how much entangled the state is. This answer is given by computing the
logarithmic negativity \cite{logneg}, so far the only computable
measure for states in infinite-dimensional Hilbert spaces.
It is defined as the sum of the symplectic eigenvalues of
$\mg^{\rm P.T.}$, which translates as \cite{Eisert02}
\begin{equation}
\label{eq:logneg}
E_N(\hat{\varrho}) =
\left\{ 
\begin{array}{cc} 
-(\log \circ f)(\mg)\,, & \mbox{if } f(\mg)<1, \\[.5ex] 
0 & \mbox{otherwise},
\end{array} \right.
\end{equation}
where
\begin{equation}
\hspace*{-6ex}
f(\mg) = \left( {\textstyle\frac{1}{2}}\left(\det\textbf{C}_1 \!+\!
\det\textbf{C}_2\right) \!-\! \det\textbf{C}_3 \!-\!
\sqrt{\left[{\textstyle\frac{1}{2}}(\det\textbf{C}_1
\!+\!\det\textbf{C}_2)\!-\!\det\textbf{C}_3\right]^2-\det\mg} \right)^{1/2}.
\end{equation}
An immediate consequence is that it is impossible to distill Gaussian
states with Gaussian operations \cite{Eisert02,NonDistill}. 
We are now in the position to look at practically important situations
such as entanglement degradation and quantum teleportation
in noisy environments.


\section{Entanglement degradation}
\label{sec:degradation}

Let us consider the influence of passive optical devices
on entangled two-mode quantum states. We have seen
in the previous section
that Gaussian states are fully characterized by their (first and)
second moments. In order to describe the influence of dielectric objects
on the quantum state we therefore need to know only how these moments
transform. The easiest way to see this is to look at the quantum-optical
input-output relations \cite{Buch,Gruner96}. They relate the amplitude
operators of the electromagnetic field impinging on the dielectric object
to the amplitude operators of the field leaving the device. In the following
we will assume that the dielectric material shows absorption as it is
generically the case. This is due to the fact that the dielectric
permittivity
(as well as the magnetic susceptibility)
has to fulfil the
Kramers--Kronig relations, which connect the real part of the permittivity
(responsible for dispersion) to the imaginary part (responsible for
absorption). That said one recognizes that quantization of the
electromagnetic field in absorbing matter has to be performed.

Here we will use the
macroscopic description \cite{Buch,Quantisierung} that
is most suitable for the situation we have in mind.
The electromagnetic field is quantized by expanding it in terms of a complete
set of bosonic vector fields $\hat{\textbf{f}}(\textbf{r},\omega)$ that
describe collective excitations of the electromagnetic field, the matter
polarization, and the reservoir responsible for absorption, viz.
\begin{equation}
\label{eq:3.1}
\hat{\textbf{E}}(\textbf{r}) = i\sqrt{\frac{\hbar}{\pi\varepsilon_0}}
\int d^3{s} \int\limits_0^\infty d\omega \, \frac{\omega^2}{c^2}\,
\sqrt{{\rm Im}\,\varepsilon(\textbf{s},\omega)}
\mbb{G}(\textbf{r},\textbf{s},\omega) \,
\hat{\textbf{f}}(\textbf{s},\omega)
+\mbox{H.c.}\,,
\end{equation}
with $\mbb{G}(\textbf{r},\textbf{s},\omega)$ being the classical
Green tensor.
Expanding the electromagnetic field in terms of amplitude operators outside
the dielectric device where no matter is present, one can derive the
input-output relation
\begin{equation}
\label{eq:3.2}
\hat{\textbf{b}}(\omega)
=\textbf{T}(\omega)
\hat{\textbf{a}}(\omega)
+\textbf{A}(\omega)
\hat{\textbf{g}}(\omega)
\end{equation}
in which the $\hat{a}_i(\omega)$ and $\hat{b}_i(\omega)$
($i$ $\!=$ $\!1,2$) denote the amplitude
operators of the incoming and outgoing fields at frequency $\omega$,
and the
$\hat{g}_i(\omega)$ denote the noise operators associated with absorption
inside the device. The matrices $\textbf{T}(\omega)$ and $\textbf{A}(\omega)$
are the transformation and absorption matrices, respectively.
They are determined by the Green tensor and satisfy the
(energy-)conservation relation 
\mbox{$\textbf{T}(\omega)\textbf{T}^+(\omega)$ $\!+$
$\!\textbf{A}(\omega)\textbf{A}^+(\omega)$ $\!=$ $\!\openone$}.
With regard to Gaussian states, we need to know how second moments of the
amplitude operators transform. Assuming that the incoming field modes and
the device are not correlated, we derive, for example,
\begin{eqnarray}
\langle \hat{b}_1^\dagger(\omega) \hat{b}_1(\omega) \rangle
\hspace{-1ex}&=&\hspace{-1ex}
|T_{11}(\omega)|^2 \langle\hat{a}_1^\dagger(\omega)\hat{a}_1(\omega)\rangle
+|T_{12}(\omega)|^2 \langle\hat{a}_2^\dagger(\omega)\hat{a}_2(\omega)\rangle
\nonumber \\
\hspace{-1ex}&&\hspace{-1ex}
+\,|A_{11}(\omega)|^2 \langle\hat{g}_1^\dagger(\omega)\hat{g}_1(\omega)\rangle
+|A_{12}(\omega)|^2 \langle\hat{g}_2^\dagger(\omega)\hat{g}_2(\omega)\rangle.
\end{eqnarray}

Let us now look at the most generic
two-mode entangled Gaussian state, the
two-mode squeezed vacuum (TMSV). In the Fock basis it reads
($\zeta\in\mathbb{R}^+$)
\begin{equation}
\label{eq:3.4}
|\mbox{TMSV}\rangle = 
\exp \zeta(\hat{a}_1\hat{a}_2-\hat{a}_1^\dagger\hat{a}_2^\dagger) |00\rangle
=\sqrt{1-q^2} \sum\limits_{n=0}^\infty q^n |nn\rangle
\,,\quad q=\tanh\zeta \,.
\end{equation}
It is separable only if $q$ $\!=$ $\!0$, otherwise it is entangled,
where the entanglement content is
$E$ $\!=$ $\!-\ln (1-q^2)$ $\!-$
$\!q^2/(1-q^2)
\ln q^2$ $\lesssim 2\zeta$ $\!=$ $\!E_N$.
In the language used in Section \ref{sec:gauss}, the TMSV translates
into a covariance matrix 
\begin{equation}
\label{eq:3.5}
\mg = \left( \begin{array}{rrrr}
c&0&s&0\\0&c&0&-s\\s&0&c&0\\0&-s&0&c
\end{array} \right),
\quad
c=\cosh 2\zeta,
\quad
s=\sinh 2\zeta. 
\end{equation}
We are interested how the entanglement changes when the
two modes are transmitted through \textit{two} fibers with equal
transmission coeffcients $|T|$, reflection coefficients $|R|$, and mean
thermal photon numbers $n_{\rm th}$. Applying the input-output
relations (\ref{eq:3.2}), the second moments transform as 
\begin{equation}
c \to c|T|^2+|R|^2+(2n_{\rm th}+1)(1-|T|^2-|R|^2)\,,\quad
s \to s|T|^2,
\end{equation}
which translates into a transformation of the covariance matrix as
\begin{equation}
\label{eq:3.6}
\mg \to |T|^2\mg + \left[ |R|^2
+(2n_{\rm th}+1)(1-|T|^2-|R|^2) \right] \openone,
\end{equation}
from which we can easily identify the matrices $\mbb{A}$ and $\mbb{G}$
in the transformation (\ref{eq:2.7}).
Note that we have not taken care about local phases
since they do not play any r\^{o}le in determining separability and
entanglement.
Applying the criterion (\ref{eq:sep}) to the covariance matrix
(\ref{eq:3.6}), we can easily rewrite the condition of separability
as \cite{ScheelPhD}
\begin{equation}
\label{eq:3.7}
n_{\rm th} \ge \frac{|T|^2(1-e^{-2\zeta})}{2(1-|R|^2-|T|^2)}\,.
\end{equation}

Restricting ourselves
to vanishing incoupling losses ($R$ $\!=$ $\!0$) and
imposing the Lambert--Beer law of extinction $|T|$ $\!=$
$\!e^{-l/l_A}$, with $l_A$ being
the characteristic absorption length of the fibers,
from Eq.~(\ref{eq:3.7})
we derive for
the fiber length after which the TMSV becomes separable
$l_S=\frac{1}{2}l_A \ln [ 1+ (1-e^{-2\zeta}) /2n_{\rm th} ]$. 
These results say that for fibers at zero temperature entanglement
will always be present (although possibly arbitrarily small).
For finite $n_{\rm th}$ the separability length $l_S$ is a function of
the squeezing parameter $\zeta$ of the input state.

Let us now look at the actual entanglement content in terms of the logarithmic
negativity. Again, we only need to insert the transformed covariance
matrix (\ref{eq:3.6}) into Eq.~(\ref{eq:logneg}). Neglecting again the
mean thermal photon numbers of the fibers ($n_{\rm th}=0$), we derive
for the logarithmic negativity
\begin{equation}
\label{eq:3.10}
E_N = -\ln\left[1-|T|^2(1-e^{-2\zeta})\right],
\end{equation}
which is independent of the reflection coefficient $R$.
For perfect transmission \mbox{($T$ $\!=$ $\!1$)},
we obtain the familiar result \mbox{$E_N$ $\!=$ $\!2\zeta$}.
Another immediate consequence of Eq.~(\ref{eq:3.10}) is that the
transmittable entanglement saturates. In the limit of infinite
initial squeezing \mbox{($\zeta$ $\!\to$ $\!\infty$)} the maximal
entanglement that can pass through fibers of given length $l$ is just
$E_{N,\max}$ $\!=$ $\!-$ $\!\ln\bigl[1$ $\!-$ $\!|T|^2\bigr]$
$\!=$ $\!-$ $\!\ln\bigl[ 1$ $\!-$ $\!e^{-2l/l_A}\bigr]$.
In other words, no matter how strong the initial squeezing was, the amount
of transmittable entanglement is limited by the fiber properties (in
this simple model the fiber length normalized to the characteristic
absorption length).
Moreover, the entanglement depends in a simple manner on the
overall losses (accounting for reflection and absorption, since
$1$ $\!-$ $\!|T|^2$ $\!=$ $\!|R|^2$ $\!+$ $\!|A|^2$) during transmission.
Figure \ref{fig:logneg} shows the dependence of the maximal
entanglement on the fiber length.
It says that in order to transmit the quantum information equivalent to
$E_N$ $\!=$ $\!1$,
the transmission distance must not exceed the value
\mbox{$l$ $\!=$ $\!l_A/2\ln 2$ $\!\approx$ $\!0.35\,l_A$}.
Equivalently, the transmission coefficient must not be smaller
than $|T|^2$ $\!=$ $\!0.5$.

Let us apply this result to quantum dense coding considered, where a
classical bit is encoded in a coherent shift of one mode of a TMSV. 
The result in \cite{Ban00} is that classical coding is superior to
quantum dense coding whenever the transmission coefficient $|T|^2$
drops below the value of $0.75$, which corresponds, according to
Eq.~(\ref{eq:3.10}), to a log-negativity of $2$. The example clearly
shows the limits of the performance of quantum information 
processes due to the unavoidably existing entanglement degradation.


\section{Quantum teleportation in noisy environments}
\label{sec:teleport}

A widely discussed process in quantum information theory has been
quantum teleportation \cite{Teleportation}. Its purpose is to transport
a substantial amount of information about an unknown (single)
quantum state spatially from one location to another
without transmitting the state itself. The idea is the
following. A signal mode that is prepared in the 
unknown quantum state is mixed with one mode of a (maximally)
entangled two-mode state, and the resulting output quantum state
is subsequently measured by homodyne detection. The (single-)measurement
result is submitted classically to the receiver who, depending on the
measurement result, performs a coherent displacement of
the quantum state in which his mode (of the two-mode entangled system)
has been prepared owing to the measurement.

In order to make contact with the theory developed in
the previous sections, we restrict
ourselves to the case in which the unknown state is a Gaussian
(with zero mean). 
That is, the initial state in mode 0 can be
described by a ($2\times 2$) covariance matrix
$\mg_{\rm in}$. This class covers all squeezed and thermal states and
combinations of them. Accordingly, the source of the entangled state
in modes 1 and 2 is assumed to produce a TMSV with a covariance matrix
$\mg_{\rm TMSV}$ as in Eq.~(\ref{eq:3.5}).
The covariance matrix of the total state
at the beginning of the teleportation process is then the direct sum
$\mg_{\rm in}\oplus\mg_{\rm TMSV}$.

As already mentioned, the transmission lines from the entanglement
source to the sender and the receiver, which are realized, e.g., by
fibers, are not perfect in practice. The question we would like to
address is to what extent the teleportation protocol has to be
modified when decoherence is present. Especially, we will try to
answer the question where the source of the entangled state has to be
located to obtain an optimal teleportation protocol.


\subsection{Imperfect teleportation}

Let us slightly generalize the situation,
compared to the previous section, in that we consider the possibility of
having two different fibers connecting the entanglement source with the
sender and the receiver. The matrix $\mg_{\rm TMSV}$ will thus change to
\cite{Scheel00c}
\begin{equation}
\mg_{\rm TMSV} \to \mg_{\rm dec} =
\left( \begin{array}{rrrr}
a  & 0  & c_1& c_2 \\
0  & a  & c_2& -c_1\\
c_1& c_2& b  & 0  \\
c_2&-c_1& 0  & b 
\end{array} \right)
\equiv \left( \begin{array}{ll} 
\textbf{A} & \textbf{C} \\ \textbf{C}^{\rm T} & \textbf{B}
\end{array} \right)
\end{equation}
with
\begin{eqnarray}
     \label{eq:a}
a &=& c|T_1|^2+|R_1|^2+(2n_{{\rm th},1}+1)(1-|T_1|^2-|R_1|^2) ,\\
      \label{eq:b}
b &=& c|T_2|^2+|R_2|^2+(2n_{{\rm th},2}+1)(1-|T_2|^2-|R_2|^2) ,
\end{eqnarray}
\begin{eqnarray}
\label{eq:c}
c_1&=& s \textrm{Re}(T_1T_2),
\\
\label{eq:c2}
c_2&=& s \textrm{Im}(T_1T_2).
\end{eqnarray}
Here, the $T_i$ and $R_i$ ($i$ $\!=$ $\!1,2$)
are the transmission and reflection coefficients
of the $i$th fiber.

The operation of mixing one mode of the transmitted TMSV
(covariance matrix $\mg_{\rm dec}$) with
the signal mode at a symmetric beam splitter
is described by a symplectic matrix of the form
\begin{equation}
\textbf{S}_{\rm BS} = \frac{1}{\sqrt{2}} \left(
\begin{array}{rr}
\openone & \openone \\ -\openone & \openone
\end{array} \right) \,.
\end{equation}
Remembering that $\textbf{S}_{\rm BS}$ acts only on modes 0 and 1,
the covariance matrix of the tripartite state reads as
$\mg_{012}$ $\!=$ $\!(\textbf{S}_{\rm BS}\oplus\openone_2)$
$(\mg_{\rm in}\oplus\mg_{\rm dec})(\textbf{S}_
{\rm BS}\oplus\openone_2)^{\rm T}$,
where $\openone_2$ is the identity operation on mode 2, the receiver's side.
In block matrix notation, $\mg_{012}$ reads
\begin{eqnarray}
\mg_{012} 
&=& \left( \begin{array}{ccc}
(\mg_{\rm in}+\textbf{A})/2 & 
(\textbf{A}-\mg_{\rm in})/2 &
\textbf{C}/\sqrt{2} \\
(\textbf{A}-\mg_{\rm in})/2 & 
(\mg_{\rm in}+\textbf{A})/2 &
\textbf{C}/\sqrt{2} \\
\textbf{C}^{\rm T}/\sqrt{2} &
\textbf{C}^{\rm T}/\sqrt{2} & \textbf{B}
\end{array} \right) 
:=\left( \begin{array}{ll}
\textbf{M} & \textbf{N} \\ \textbf{N}^{\rm T} & \textbf{B}
\end{array} \right).
\end{eqnarray}

As shown in the Appendix, homodyne detection
(with respect to the variables $x_0$ and $p_1$ after the beam splitter)
is equivalent to a partial Fourier transform
[Eqs.~(\ref{app:1}) and (\ref{app:4})].
Let us denote the measurement outcomes corresponding to the
variables $x_0$ and $p_1$ by $X_{x_0}$ and $X_{p_1}$, respectively.
The characteristic function at the receiver's side will then be
\begin{equation}
\label{eq:cfrec}
\hspace*{-3ex}
\chi_{\rm rec}(\mbb{\lambda}_2) = p(X_{x_0},X_{p_1})
\exp \left[ -\frac{1}{4} \mbb{\lambda}_2^{\rm T} \mg_{\rm rec}
\mbb{\lambda}_2 + i\, \mbb{\lambda}_2^{\rm T} 
\textbf{N}^{\rm T} (\mbb{\pi}\textbf{M}\mbb{\pi})^{\rm MP}
\left(\! \begin{array}{r}
X_{x_0} \\ -X_{p_1}
\end{array} \!\right)
\right].
\end{equation}
The probability of obtaining a certain pair of measurement outcomes is
\begin{equation}
\hspace*{-3ex}
p(X_{x_0},X_{p_1}) = 
\frac{1}{\pi\sqrt{\det(\mbb{\pi}\textbf{M}\mbb{\pi})}}
\exp\!\left[ -
\begin{array}{cc} ( \, X_{x_0} & \hspace{-1ex}-X_{p_1} \, ) \\ & \end{array} 
(\mbb{\pi}\textbf{M}\mbb{\pi})^{\rm MP}
\left(\! \begin{array}{r}
X_{x_0} \\ -X_{p_1}
\end{array} \!\right)
\right] ,
\end{equation}
and the covariance matrix of the receiver's quantum state obtains the form
\begin{equation}
\label{eq:schur}
\mg_{\rm rec} =
\textbf{B}-\textbf{N}^{\rm T} (\mbb{\pi}\textbf{M}\mbb{\pi})^{\rm MP}\textbf{N},
\end{equation}
where $\mbb{\pi}$ $\!=$ $\!\mbox{diag}(0,1,1,0)$ and the
superscript MP denotes the
Moore--Penrose inverse. The matrix $\mbb{\pi}$ is, in fact, a projector
onto the $(p_0,x_1)$-plane. It removes all entries of $\textbf{M}$ except
the square block matrix specified by the $1$-entries. The Moore--Penrose
inverse of a matrix containing a square block on the diagonal and zeros
otherwise is nothing but a matrix with the inverse of the
square block at the same position (and zero everywhere else).

If we denote the elements of the covariance matrix of the unknown
signal state by
$\mg_{\rm in}$ $\!=$ $\!\left(\begin{array}{cc}x&z\\z&y\end{array} \right)$,
the involved matrices read
\begin{equation}
\label{eq:matrizen}
\hspace*{-3ex}
\textbf{N} = \left( \begin{array}{rr}
c_2 & -c_1 \\ c_1 & c_2
\end{array} \right) , \quad
(\mbb{\pi}\textbf{M}\mbb{\pi})^{\rm MP} = \frac{1}{(a+x)(a+y)-z^2}
\left( \begin{array}{cc}
a+x & z \\ z & a+y
\end{array} \right) ,
\end{equation}
and the covariance matrix (\ref{eq:schur}) thus takes the form of
\begin{eqnarray}
\label{eq:4.10}
\lefteqn{
\mg_{\rm rec} = \left( \begin{array}{cc}
b & 0 \\ 0 & b
\end{array} \right)
-\frac{1}{(x+a)(y+a)-z^2}
} \nonumber \\[.5ex] &&\hspace{-1.75ex}
\times
\left(\!\! \begin{array}{cc}
c_2^2(x+a)+c_1^2(y+a)+2c_1c_2z & c_1c_2(y-x)-z(c_1^2-c_2^2) \\
c_1c_2(y-x)-z(c_1^2-c_2^2) & c_1^2(x+a)+c_2^2(y+a)-2c_1c_2z
\end{array} \!\!\right)\!.
\end{eqnarray}
In the limit $\zeta\to\infty$, perfect transmission of the TMSV,
and the phase adjusted such that $c_2$ $\!=$ $\!0$,
the covariance matrix on the receiver's side becomes
$\mg_{\rm rec}$ $\!=$ $\!\mg_{\rm in}$.
In particular,
we recover that for perfect teleportation an infinitely
squeezed TMSV (with $\zeta\to\infty$) is needed. 
Note that $\mg_{\rm rec}$ does not depend on the measurement
result, even for non-perfect transmission. That means that the information
about the quantum features of a Gaussian state are not transported via the
classical channel. What we see, though, is that the covariance matrix is
sensitive to the properties of the channel itself.
\begin{figure}[h]
\begin{minipage}[t]{0.45\textwidth}
\includegraphics[width=\textwidth]{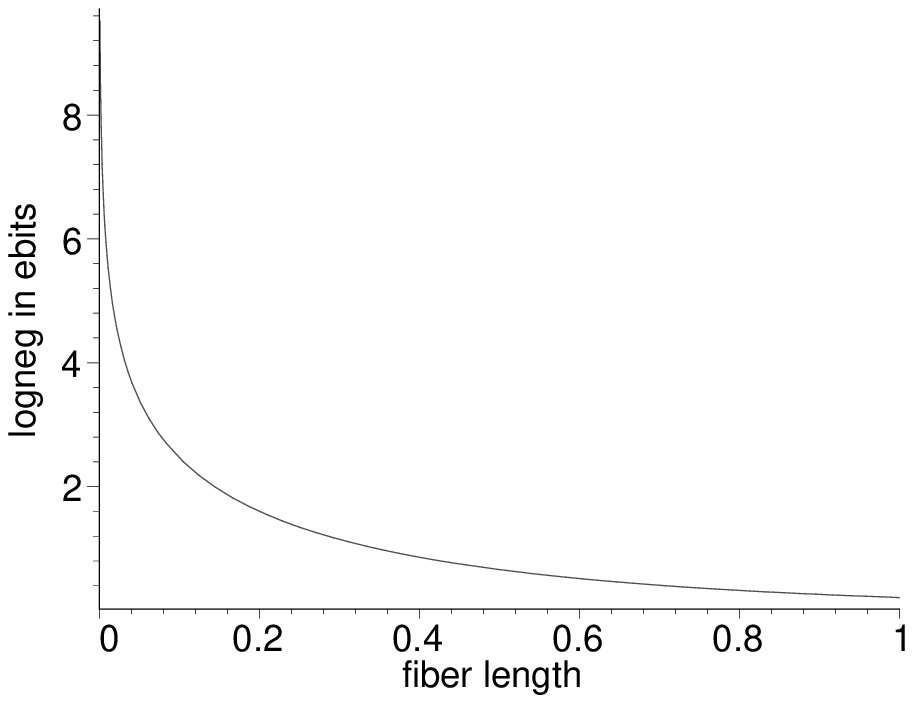}
\caption{\label{fig:logneg} Maximally transmittable entanglement 
as a function of the fiber length (in units of the absorption length) 
for the case of a TMSV input.}
\end{minipage}\hfill
\begin{minipage}[t]{0.5\textwidth}
\includegraphics[width=\textwidth]{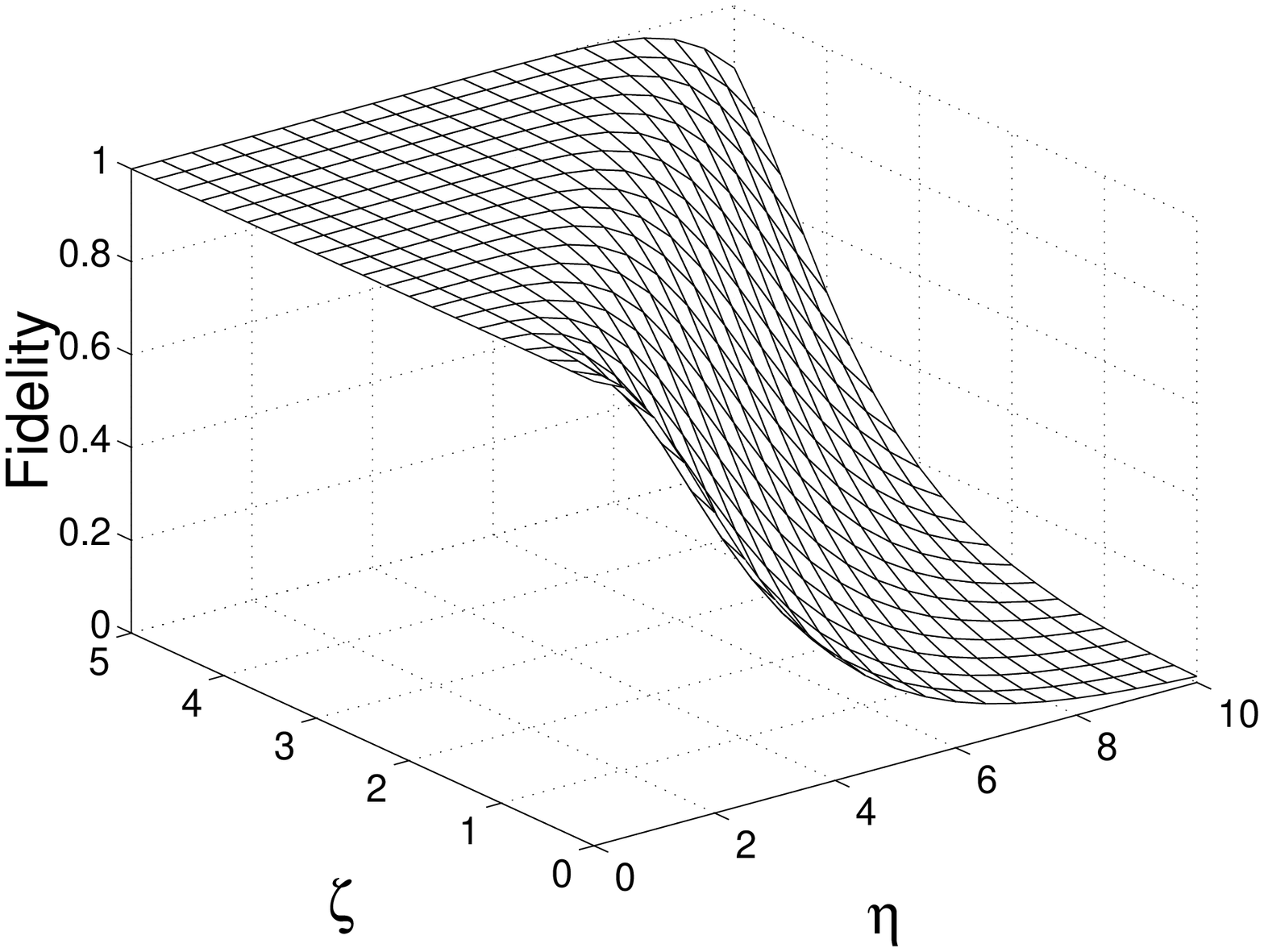}
\caption{\label{fig:purefid} Teleportation fidelity
$F_{\rm qu}$ of a (pure) squeezed state (squeezing parameter $\eta$)
by means of an unperturbed TMSV (squeezing parameter $\zeta$).
}
\end{minipage}
\end{figure}

\subsection{Teleportation fidelity}

Generically, the TMSV is neither infinitely squeezed nor are the
transmission lines from the entanglement source to sender and receiver
perfect. Hence, one needs a measure for the success rate of the
teleportation protocol. For teleporting pure quantum states, one
commonly uses the fidelity defined as the overlap between the
quantum state to be teleported and the quantum state at the receiver.
In terms of their characteristic functions, the fidelity is given as a
double integral over the (complex) parameter $\mbb{\lambda}$ as
\begin{equation}
\label{eq:4.11}
F = \frac{1}{\pi}\int d^2\mbb{\lambda} \, \chi_{\rm in}(\mbb{\lambda})
\chi_{\rm rec}(-\mbb{\lambda}) .
\end{equation}

For Gaussian states with zero mean the negative argument in
Eq.~(\ref{eq:4.11}) does not matter since the characteristic function
is an exponential quadratic form of $\mbb{\lambda}$.
As we have seen, the homodyne detection introduces a coherent
displacement, but does not influence the covariance matrix.
With regard to the quantum properties, it is therefore
sufficient to compute the overlap between
characteristic functions with zero mean.
In that way, from Eq.~(\ref{eq:4.11}) we find that  
\begin{equation}
\label{eq:fidelity}
F \to
F_{\rm qu} = \frac{2}{\sqrt{\det(\mg_{\rm in}+\mg_{\rm rec})}} \,.
\end{equation}

For perfect teleportation, the covariance matrix of the receiver's
state is identical to the covariance matrix of the signal state to be
teleported, and the fidelity becomes
$F_{\rm qu}$ $\!=$ $\!(\det\mg_{\rm in})^{-1/2}$ $\!=$ $\!1$.
Even if entanglement degradation be disregarded, perfect
teleportation cannot be realized, because of finite squeezing. Let us
consider a (pure) squeezed signal state and disregard entanglement
degradation, so that the entangled state is simply the unchanged
(pure) TMSV state. 
The parameters to look at are
$a$ $\!=$ $\!b$ $\!=$ $\!c$ $\!=$ $\!\cosh 2\zeta$,
$c_1$ $\!=$ $\!s$ $\!=$ $\!\sinh 2\zeta$,
\mbox{$c_2$ $\!=$ $\!0$}, $x$ $\!=$ $\!y$ $\!=$ $\!\cosh\eta$, $z=\sinh\eta$.
The teleportation fidelity according to Eq.~(\ref{eq:fidelity}) is then
\begin{equation}
F_{\rm qu} = \sqrt{1-\frac{\sinh^2\eta}{(\cosh\eta+\cosh 2\zeta)^2}}
\equiv \sqrt{1-\frac{\sinh^2\eta}{(\cosh\eta+\cosh E_N)^2}}
\end{equation}
[cf. Eq.~(\ref{eq:3.10})], 
whose behaviour is illustrated in Fig.~\ref{fig:purefid}.
We see that for chosen squeezing parameter $\eta$ of the
signal state the fidelity is a monotonous function of the inserted
entanglement $E_N$.
In particular, the classical teleportation fidelity
(obtained for $\zeta$ $\!=$ $\!0$) is read off as $\sqrt{2/(1+\cosh\eta)}$.


\subsection{Choice of the coherent displacement}

There have been some discussions recently about what kind of coherent
displacement should be applied \cite{WelschBoston,Discussion}. Let us
look at Eq.~(\ref{eq:cfrec}). In the limit of an infinitely 
entangled TMSV the product
$\textbf{N}^{\rm T}(\mbb{\pi}\textbf{M}\mbb{\pi})^{\rm MP}$ 
becomes the symplectic matrix $\mbb{\Sigma}$ which is independent of the
initial state. Thus, in the ideal case the coherent displacement is solely
determined by the measurement results. This is the standard
teleportation protocol, where the coherent displacement
to be performed on the
receivers's side is exactly \mbox{$X_{x_0}$ $\!-$ $\!iX_{p_1}$}.

Even if an infinitely entangled TMSV were available, entanglement
degradation would prevent one from observing this simple result.
On recalling Eq.~(\ref{eq:matrizen}) together with
Eqs.~(\ref{eq:a}) and (\ref{eq:c}),
the relevant matrix product
$\textbf{N}^{\rm T}(\mbb{\pi}\textbf{M}\mbb{\pi})^{\rm MP}$ now reads
\begin{equation}
\label{eq:shift2}
\lim_{\zeta\to\infty}
\textbf{N}^{\rm T}(\mbb{\pi}\textbf{M}\mbb{\pi})^{\rm MP}
=
\frac{1}{|T_1|^2}
\left( \begin{array}{rr}
\textrm{Im}\left(T_1T_2\right) & \textrm{Re}\left(T_1T_2\right) \\
\textrm{Re}\left(T_1T_2\right) & -\textrm{Im}\left(T_1T_2\right) 
\end{array} \right) .
\end{equation}
If we write the transmission coefficients as
$T_i$ $\!=$ $\!|T_i|e^{i\varphi_i}$, Eq.~(\ref{eq:shift2}) becomes
\begin{eqnarray}
\label{eq:displacement}
\lim_{\zeta\to\infty}
\textbf{N}^{\rm T}(\mbb{\pi}\textbf{M}\mbb{\pi})^{\rm MP}
&=& \mbb{\Sigma} \left| \frac{T_2}{T_1} \right| 
\left( \begin{array}{rr}
\cos\left(\varphi_1\!+\!\varphi_2\right)
& -\sin\left(\varphi_1\!+\!\varphi_2\right) \\
\sin\left(\varphi_1\!+\!\varphi_2\right)
& \cos\left(\varphi_1\!+\!\varphi_2\right)
\end{array} \right) .
\end{eqnarray}
We see that, apart from the (irrelevant) phase shift
$-(\varphi_1$ $\!+$ $\!\varphi_2)$, the standard teleportation
protocol must be modified in so far that the absolute value
of the coherent displacement has to be scaled by $|T_2/T_1|$.
This result is in agreement with
\cite{WelschBoston}. Note that the displacement (in this limit) is still
independent of the unknown state and no averaging has to be performed. 
We like to emphasize that this limit is actually the only situation in
which the word `teleportation' makes sense.

In practice, an infinitely entangled state is not available.
Nevertheless, one might want to apply the displacement
(\ref{eq:displacement}), which is optimal for infinite
entanglement. This, however, introduces further loss in fidelity.
The modification is an exponential function that depends on the
displacement of the signal state and the state at the receiver's side
after performing the displacement.

These results show that understanding decoherence and its associated
processes is important for quantum information processing in the
presence of absorption.

\section*{Acknowledgements}

We would like to thank K.~Audenaert, A.V.~Chizhov,
J.~Eisert, P.L.~Knight, L.~Kn\"oll, M.B.~Plenio, and W.~Vogel for many
fruitful discussions on the subject. 
This work has been funded in parts by a Feodor-Lynen fellowship of the
Alexander von Humboldt foundation (S.~Scheel), the QUEST programme of
the European Union (HPRN-CT-2000-00121), and the Deutsche
Forschungsgemeinschaft (DFG-Schwerpunkt Quanteninformationsverarbeitung).


\section*{Appendix: Homodyne detection}
\label{app:homodyne}
\setcounter{equation}{0}
\renewcommand{\theequation}{A.\arabic{equation}}

Perfect homodyne detection is a projective quadrature component
measurement. Thus, we have
\begin{equation}
\label{app:1}
{}_2\langle X,\varphi | \hat{\varrho} | X,\varphi \rangle_2 =
\frac{1}{\pi^2} \int d^2\mbb{\lambda}_1 d^2\mbb{\lambda}_2 \,
\hat{D}^\dagger(\mbb{\lambda}_1) \chi(\mbb{\lambda}_1,\mbb{\lambda}_2)
\,{}_2\langle X,\varphi | \hat{D}^\dagger(\mbb{\lambda}_2)
| X,\varphi \rangle_2,
\end{equation}
where the $|X,\varphi\rangle$ are the quadrature-component eigenstates
with eigenvalues $X$ for chosen
phase parameter $\varphi$ ($\varphi$ $\!=$ $\!0$ corresponds to
an $x$ measurement, $\varphi$ $\!=$ $\!\pi/2$ to a $p$ measurement).
It can be expanded as
\cite{vogelwelsch}
\begin{equation}
|X,\varphi\rangle =
\pi^{-1/4} e^{-X^2/2}
\exp\! \left[ 
-{\textstyle\frac{1}{2}}
\left( \hat{a}^\dagger e^{i\varphi} \right)^2 + 2^{1/2} X
\hat{a}^\dagger e^{i\varphi}  \right] | 0 \rangle ,
\end{equation}
and hence the diagonal-matrix elements of the coherent displacement
operator read
\begin{equation}
\label{app:3}
{}_2\langle X,\varphi | \hat{D}^\dagger(\mbb{\lambda}_2)
| X,\varphi \rangle_2 =
\delta(x_2\cos\varphi+p_2\sin\varphi)
\,\exp\left[ iX \left( x_2\sin\varphi
-p_2\cos\varphi \right) \right]
\end{equation}
[$\mbb{\lambda}_2^{\rm T}=(x_2,p_2)/\sqrt{2}$].
For $\varphi$ $\!=$ $\!0,\pi/2$, Eq.~(\ref{app:3}) reduces to
\begin{equation}
\label{app:4}
{}_2\langle X,\varphi | \hat{D}^\dagger(\mbb{\lambda}_2)
| X,\varphi \rangle_2 =
\left\{ \begin{array}{ll}
\delta(x_2)e^{-iXp_2}  & (\varphi=0),
\\
\delta(p_2)e^{iXx_2}  & 
(\varphi=\pi/2). \end{array} \right. 
\end{equation}
Equivalently, one can use the relation for the projector onto a
quadrature-component eigenstate in the form \cite{vogelwelsch}
\begin{equation}
| X,\varphi \rangle \langle X,\varphi | = \frac{1}{2\pi} \int dy\,
e^{-iyX} \hat{D}\!\left( \frac{iye^{i\varphi}}{\sqrt{2}} \right)
\end{equation}
and use the completeness of the quadrature-component states
$|X,\varphi\rangle$ and the relation
$\mbox{Tr}[\hat{D}(\lambda)\hat{D}(\lambda')]$ $\!=$
$\!\pi\delta(\lambda+\lambda')$.
On the level of the covariance matrices, the $\delta$ function leads to
the projection matrix $\mbb{\pi}$ in Eq.~(\ref{eq:schur}) since it
deletes all matrix elements associated with the measured
variables. Performing the remaining Gaussian integration, which is
equivalent to a Fourier transform, introduces the Schur complement
with respect to the leading submatrix corresponding to the subsystem
$1$.



\begin{thebibliography}[1]{99}
\bibitem{Arvind95}
Arvind, B.~Dutta, N.~Mukunda, and R.~Simon, \textit{quant-ph/9509002}.
\bibitem{Titulaer65}
U.M.~Titulaer and R.J.~Glauber, Phys. Rev. \textbf{140}, B676 (1965).
\bibitem{Vogel00}
W.~Vogel, Phys. Rev. Lett. \textbf{84}, 1849 (2000).
\bibitem{Marian93}
P.~Marian and T.A.~Marian, Phys. Rev. A \textbf{47}, 4474 (1993); 4487 (1993).
\bibitem{Scheel01}
S.~Scheel and D.-G.~Welsch, Phys. Rev. A \textbf{64}, 063811 (2001).
\bibitem{Marian02}
P.~Marian, T.A.~Marian, and H.~Scutaru, Phys. Rev. Lett. 
\textbf{88}, 153601 (2002).
\bibitem{Naimark}
M.A.~Naimark, C.R. Acad. Sci. USSR \textbf{41}, 359 (1943).
\bibitem{Eisert02}
J.~Eisert, S.~Scheel, and M.B.~Plenio, \textit{quant-ph/0204052},
accepted for publication in Phys. Rev. Lett.
\bibitem{Horn}
R.A.~Horn and C.R.~Johnson, \textit{Matrix Analysis} (Cambridge
University Press, Cambridge, 1985).
\bibitem{Duan00}
L.M.~Duan, G.~Giedke, J.I.~Cirac, and P.~Zoller, Phys. Rev. Lett. 
\textbf{84}, 2722 (2000).
\bibitem{Simon00}
 R.~Simon, Phys. Rev. Lett. \textbf{84}, 2726 (2000).
\bibitem{PeresHorodecki}
A.~Peres, Phys. Rev. Lett. \textbf{77}, 1413 (1996);
P.~Horodecki, Phys. Lett. A \textbf{232}, 333 (1997).
\bibitem{logneg}
J.~Eisert, PhD thesis (Potsdam, February 2001);
G.~Vidal and R.F.~Werner, \textit{quant-ph/0102117}.
\bibitem{NonDistill}
J.~Fiur\'{a}\v{s}ek, \textit{quant-ph/0204069};
G.~Giedke and J.I.~Cirac, \textit{quant-ph/0204085}.
\bibitem{Buch}
L.~Kn\"oll, S.~Scheel, and D.-G.~Welsch, in 
\textit{Coherence and Statistics of Photons and Atoms}, ed. J.~Pe\v{r}ina
(Wiley, New York, 2001).
\bibitem{Gruner96}
T.~Gruner and D.-G.~Welsch, Phys. Rev. A \textbf{54}, 1661 (1996).
\bibitem{Quantisierung}
T.~Gruner and D.-G.~Welsch, Phys. Rev. A \textbf{53}, 1818 (1996);
Ho~Trung~Dung, L.~Kn\"oll, and D.-G.~Welsch, Phys. Rev. A \textbf{57},
3931 (1998);
S.~Scheel, L.~Kn\"oll, and D.-G.~Welsch, Phys. Rev. A \textbf{58}, 700 (1998).
\bibitem{ScheelPhD}
S.~Scheel, PhD thesis (Jena, July 2001).
\bibitem{Ban00}
M.~Ban, J. Opt. B: Quantum Semiclass. Opt. \textbf{2}, 786 (2000).
\bibitem{Scheel00c}
S.~Scheel, L.~Kn\"oll, T.~Opatrn\'{y}, and D.-G.~Welsch, Phys. Rev. A
\textbf{62}, 043803 (2000).
\bibitem{Teleportation}
C.H.~Bennett
et al.,
Phys. Rev. Lett. \textbf{70}, 1895 (1993);
S.L.~Braunstein and H.J.~Kimble, Phys. Rev. Lett. \textbf{80}, 869 (1998).
\bibitem{WelschBoston}
D.-G.~Welsch, S.~Scheel, and A.V.~Chizhov, Proceedings of the 7th
International Conference on Squeezed States and Uncertainty Relations,
Boston, 2001 (www.wam.umd.edu/$\sim$ys/boston.html).
\bibitem{Discussion}
M.S.~Kim, Phys. Rev. A \textbf{64}, 012309 (2001);
A.V.~Chizhov, L.~Kn\"oll, and D.-G.~Welsch, Phys. Rev. A \textbf{65}, 
022310 (2002).
\bibitem{vogelwelsch}
W.~Vogel, S.~Wallentowitz, and D.-G.~Welsch, 
\textit{Quantum Optics, An Introduction} (Wiley-VCH, Berlin, 2001).
\end{thebibliography}
\end{document}